\documentclass[12pt]{article}
\usepackage{amssymb,amsmath,epsfig}
\allowdisplaybreaks

\begin{document}

\title{\bf Extended Black Hole Solutions in Self-interacting Brans-Dicke Theory}
\author{M. Sharif \thanks{msharif.math@pu.edu.pk} and Amal Majid
\thanks{amalmajid89@gmail.com}\\
Department of Mathematics, University of the Punjab,\\
Quaid-e-Azam Campus, Lahore-54590, Pakistan.}

\date{}

\maketitle
\begin{abstract}
In this paper, we formulate black hole solutions through extended
gravitational decoupling scheme in the framework of self-interacting
Brans-Dicke theory. The addition of a new source in the matter
distribution increases the degrees of freedom in the system of field
equations. Transformations in radial as well as temporal metric
functions split the system into two arrays. Each array includes the
effects of only one source (either seed or additional). The seed
source is assumed to be a vacuum and the corresponding system is
specified through the Schwarzschild metric. In order to construct a
suitable solution of the second system, constraints are applied on
the metric potentials and energy-momentum tensor of the additional
source. We obtain three solutions corresponding to different values
of the decoupling parameter in the presence of a massive scalar
field. The extra source is classified as normal or exotic through
energy conditions. It is found that two solutions agree with the
energy bounds and thus have normal matter as their source.
\end{abstract}

\section{Introduction}

The universe is a complex structure consisting of mysterious forces
and unknown matter components. In order to gain insight into the
evolution and mechanism of the cosmos, it is essential to study the
arrangement and physical features of astrophysical objects. Black
hole (BH) is one of the self-gravitating systems with a singularity
hidden behind the event horizon. These cosmic objects have a strong
gravitational field and serve as excellent laboratories to test
relativistic theories in the strong-field regime. The existence of
BHs has been strengthened due to the recent detection of
gravitational waves \cite{1}. Moreover, the study of BH shadow
\cite{2} has contributed to the significance of BH solutions with
well-behaved physical characteristics.

Schwarzschild obtained the first BH solution for a vacuum spacetime
\cite{3}. General relativity (GR) formulates surprisingly simple
solutions for BHs in accordance with the no-hair conjecture (BH
solutions cannot carry additional charges \cite{4}) with three
prominent features: mass, charge and angular momentum \cite{5}. The
validity of no-hair theorem is now being tested with improved
studies and observations of BH systems. In fact, different setups
have been constructed to evade the no-hair theorem \cite{6}. Recent
studies suggest that BHs, as sources of extreme gravity, can possess
soft quantum hair \cite{7}. However, the derivation of new solutions
representing BHs is hindered by the non-linearity as well as high
degree of freedom in the field equations.

Ovalle \cite{10} devised the method of gravitational decoupling to
obtain new models representing relativistic objects with
well-determined characteristics. In this scheme, known isotropic
solutions (or seed solutions) are extended to their anisotropic
versions by involving an additional source in the seed solution. The
array of field equations is divided into two systems through a
minimal geometric deformation (MGD), i.e., a transformation in the
radial metric component. A viable solution represents the set
corresponding to the original source whereas physically suitable
constraints are applied to formulate a solution associated with the
additional source. Combining both solutions yields the required
anisotropic extension of the known solution. The key feature of this
technique is that the seed and extra sources must be conserved
individually as they interact gravitationally only. Ovalle first
used the MGD approach to construct spherically symmetric solutions
in the framework of Randall-Sundrum braneworld \cite{10}.

The procedure of MGD was adopted to analyze the compactness of
self-gravitating objects by computing the analogue of Tolman IV in
braneworld \cite{11}. Ovalle et al. \cite{13} followed this method
to evaluate three anisotropic extensions of Tolman IV in GR.
Different anisotropic BH solutions were obtained by applying this
approach to a vacuum Schwarzschild solution \cite{14}. Gabbanelli et
al. \cite{15} evaluated anisotropic extension of the
Durgapal-Fuloria solution and discussed its viability. Estrada and
Tello-Ortiz \cite{16} formulated two anisotropic solutions by
extending Heintzmann solution. Sharif and Sadiq \cite{17}
investigated the effects of electromagnetic field on extended
Krori-Barua solution. P\'{e}rez Graterol \cite{18} formulated new
well-behaved anisotropic solutions by introducing deformations in
the radial metric component of Buchdahl spacetime. Sharif and
Ama-Tul-Mughani \cite{19} obtained anisotropic extensions of string
cloud. In a recent paper \cite{20}, Tello-Ortiz et al. built
anisotropic compact models by combining the techniques of embedding
class-one and MGD.

The approach of gravitational decoupling through MGD has proved
highly beneficial in obtaining physically relevant anisotropic
solutions. However, this method applies to the scenario in which the
considered sources do not exchange energy. Casadio et al. \cite{21}
overcame this drawback by deforming temporal as well as radial
components of the spacetime. However, this modification of MGD
approach gives valid results in the case of vacuum only and fails to
satisfy the conservation law related to astrophysical objects filled
with fluid. Recently, Ovalle \cite{22} introduced decoupling through
deforming both (radial/temporal) metric potentials. Decoupling
through extended geometric deformation (EGD) disintegrates the
system without restricting the type of matter distribution.
Contreras and Bargue\~{n}o \cite{23} applied this method to vacuum
BTZ solution in 2+1-dimensions. The EGD scheme has also been applied
to evaluate anisotropic versions of Tolman IV \cite{24} and
Krori-Barua \cite{25} solutions. Recently, Ovalle et al. \cite{26}
formulated hairy BHs by extending Schwarzschild spacetime through
the EGD technique. There are also some attempts \cite{27} to obtain
anisotropic solutions in modified theories through MGD as well as
EGD schemes.

Brans and Dicke in 1961 \cite{28} proposed a theory based on Dirac
hypothesis (gravitational constant $(G)$ is not a constant) as well
as Mach principle (inertia arises from the acceleration of matter
distribution) and introduced a long-range massless scalar field
($\psi$) that acts as the inverse of dynamical gravitational
constant. A tunable parameter $\omega_{BD}$ minimally couples the
scalar field to fluid distribution and is termed as the coupling
constant. The role of this dimensionless parameter is reduced during
the inflationary era of the universe leading to smaller values of
$\omega_{BD}$ during this phase \cite{29}. Moreover, the weak-field
tests are satisfied for $\omega_{BD}>40,000$ \cite{30} which results
in a conflict to determine the correct range of the coupling
parameter. For this purpose, the Brans-Dicke (BD) theory is
remodeled to more sophisticated versions of itself such as
self-interacting BD (SBD) theory. This theory involves a massive
scalar field ($\Upsilon$) and a potential function $V(\Upsilon)$
that re-adjusts the values of $\omega_{BD}$ \cite{31}. The coupling
parameter can assume values greater than $-\frac{3}{2}$ when the
scalar field is more massive than $2\times10^{-25}GeV$ \cite{32}.

Researchers have derived different solutions in BD theory as well as
its modifications to discuss different astrophysical phenomena.
Thorne and Dykla \cite{33} studied BHs in three dimensions and
concluded that 3-dimensional BD BHs are identical to their
counterparts in GR. Johnson \cite{34} investigated the conditions
under which static BHs in BD gravity reduce to Schwarzschild
solution. Hawking \cite{35} showed that a stationary BH metric
satisfies the BD field equations if and only if it is also a
solution of GR. Geroch method \cite{36} was employed by Sneddon and
McIntosh \cite{37} to discuss vacuum models. Bruckman and Kazes
\cite{38} considered a perfect matter source with a linear equation
of state (EoS) to formulate solutions for a spherically symmetric
spacetime. The BD field equations were converted to Einstein field
equations by Goswami \cite{39} to evaluate vacuum solutions. Riazi
and Askari \cite{40} used numerical techniques to approximate
solution for an empty sphere and studied the trend of rotation
curves. Kim \cite{41} studied thermodynamics of BHs in BD gravity to
discriminate between trivial and non-trivial BHs. Campanelli and
Lousto \cite{42} studied a family of BD solutions and determined the
range of parameters yielding BH solutions different from GR.
Recently, we have used MGD and EGD techniques to generate
astrophysical and cosmological solutions, respectively in the
context of SBD theory \cite{43}.

In this paper, we derive BH solutions in SBD theory by implementing
EGD method on Schwarzschild spacetime. The additional gravitational
source is decoupled from the original source (vacuum) by means of
deformations in the metric components. The new black hole solution
is derived for three cases of linear EoS by considering the extra
source as a tensor-vacuum. Out of these solutions, two obey the
energy conditions and are viable. We have also checked the
asymptotic behavior of the extended solutions. The paper is
organized as follows. The field equations incorporating an
additional source in matter distribution are formulated in section
\textbf{2}. In section \textbf{3}, deformations on the metric
components are applied to decouple the field equations. The physical
characteristics of BH solutions are studied in section \textbf{4}.
In the last section, a summary of the main results is presented.

\section{Self-interacting Brans-Dicke Theory}

In this section, we formulate the SBD field equations by adding an
extra source ($\Theta_{\gamma\delta}$) in the action (in
relativistic units) as
\begin{equation}\nonumber
S=\int\sqrt{-g}(\mathcal{R}\Upsilon-\frac{\omega_{BD}}{\Upsilon}\nabla^{\gamma}\nabla_{\gamma}\Upsilon
-V(\Upsilon)+\emph{L}_m+\alpha\emph{L}_\Theta)d^{4}x,
\end{equation}
where the Ricci scalar is denoted by $\mathcal{R}$ whereas the
Lagrangian densities of matter and new source are represented by
$\emph{L}_m$ and $\emph{L}_\Theta$, respectively. The extra source
may include scalar, vector or tensor fields which are generally
responsible for inducing anisotropy in the fluid distribution. The
dimensionless parameter $\alpha$ tracks the strength of the coupling
between the two matter sources $T_{\gamma\delta}^{(m)}$ and
$\Theta_{\gamma\delta}$. The SBD action provides the following field
and wave equations
\begin{eqnarray}\label{1}
G_{\gamma\delta}&=&T^{\text{(eff)}}_{\gamma\delta}+\frac{\alpha}{\Upsilon}\Theta_{\gamma\delta}
=\frac{1}{\Upsilon}(T_{\gamma\delta}^{(m)}
+T_{\gamma\delta}^\Upsilon+\alpha\Theta_{\gamma\delta}),\\\label{2}
\Box\Upsilon&=&\frac{g^{\gamma\delta}(\alpha\Theta_{\gamma\delta}+T_{\gamma\delta}^{(m)})}{3+2\omega_{BD}}+\frac{1}{3+2\omega_{BD}}
(\Upsilon\frac{dV(\Upsilon)}{d\Upsilon}-2V(\Upsilon)),
\end{eqnarray}
where $\Box$ denotes the d'Alembertian operator. We consider perfect
fluid as the seed source described by the following energy-momentum
tensor
\begin{equation}\label{3}
T_{\gamma\delta}^{(m)}=(p+\rho)u_{\gamma}u_{\delta}-pg_{\gamma\delta},
\end{equation}
where $p$, $\rho$ and $u_{\gamma}$ indicate isotropic pressure,
energy density and four velocity, respectively. The energy-momentum
tensor associated with the massive scalar field is expressed as
\begin{equation}\label{4}
T_{\gamma\delta}^\Upsilon=\Upsilon_{,\gamma;\delta}-g_{\gamma\delta}\Box\Upsilon+\frac{\omega_{BD}}{\Upsilon}
(\Upsilon_{,\gamma}\Upsilon_{,\delta}
-\frac{g_{\gamma\delta}\Upsilon_{,\alpha}\Upsilon^{,\alpha}}{2})-\frac{V(\Upsilon)g_{\gamma\delta}}{2}.
\end{equation}

The metric for a static spherically symmetric configuration is
written as
\begin{equation}\label{5}
ds^2=e^{\mu(r)}dt^2-e^{\varepsilon(r)}dr^2-r^2(d\theta^2+\sin^2\theta
d\phi^2).
\end{equation}
The field equations linking geometry of the system to matter
distribution are formulated via Eqs.(\ref{1})-(\ref{5}) as
\begin{eqnarray}\label{6}
\frac{1}{r^2}-e^{-\varepsilon}\left(\frac{1}{r^2}-\frac{\varepsilon'}{r}\right)&=&
\frac{1}{\Upsilon}(\rho+\alpha\Theta_0^0+T_0^{0\Upsilon}),\\\label{7}
-\frac{1}{r^2}+e^{-\varepsilon}\left(\frac{1}{r^2}+\frac{\mu'}{r}\right)&=&
\frac{1}{\Upsilon}(p-\alpha\Theta_1^1-T_1^{1\Upsilon}),\\\label{8}
\frac{e^{-\varepsilon}}{4}\left(2\mu''+\mu'^2-\varepsilon'\mu'+2\frac{\mu'-\varepsilon'}{r}\right)&=&
\frac{1}{\Upsilon}(p-\alpha\Theta_2^2-T_2^{2\Upsilon}),
\end{eqnarray}
where
\begin{eqnarray}\nonumber
T_0^{0\Upsilon}&=&e^{-\varepsilon}\left[\Upsilon''+\left(\frac{2}{r}-\frac{\varepsilon'}{2}
\right)\Upsilon'+\frac{\omega_{BD}}{2\Upsilon}\Upsilon'^2-e^\varepsilon\frac{V(\Upsilon)}
{2}\right],\\\nonumber
T_1^{1\Upsilon}&=&e^{-\varepsilon}\left[\left(\frac{2}{r}+\frac{\mu'}
{2}\right)\Upsilon'-\frac{\omega_{BD}}{2\Upsilon}\Upsilon'^2-e^\varepsilon\frac{V(\Upsilon)}{2})\right],\\\nonumber
T_2^{2\Upsilon}&=&e^{-\varepsilon}\left[\Upsilon''+\left(\frac{1}{r}-\frac{\varepsilon'}
{2}+\frac{\mu'}{2}\right)\Upsilon'+\frac{\omega_{BD}}{2\Upsilon}\Upsilon'^2-e^\varepsilon\frac{V(\Upsilon)}{2}
\right].
\end{eqnarray}
Here prime represents derivative with respect to the radial
coordinate. The evolution of the scalar field in Eq.(\ref{2}) for a
spherical system is evaluated as
\begin{eqnarray}\label{7}
\Box\Upsilon&=&-e^{-\varepsilon}\left[\left(\frac{2}{r}-\frac{\varepsilon'}
{2} +\frac{\mu'}{2}\right)\Upsilon'+\Upsilon''\right].
\end{eqnarray}
An inspection of the system (\ref{6})-(\ref{8}) reveals that
anisotropy of the static source does not vanish unless
$\Theta_1^1=\Theta_2^2$. Thus, the additional source plays a crucial
role in extending isotropic solutions to the domain of anisotropic
solutions.

\section{Gravitational Decoupling}

The four differential equations (\ref{6})-(\ref{7}) give rise to a
system containing nine unknowns: five matter variables
($p,~\rho,~\Theta_{0}^{0},~\Theta_{1}^{1},~\Theta_{2}^{2}$); two
metric potentials ($\mu,~\varepsilon$); massive scalar field and a
potential function. The novel approach of EGD \cite{22} reduces the
degrees of freedom in the underdetermined system by transforming the
temporal as well as radial metric potentials. These deformations are
controlled by the parameter $\alpha$ as
\begin{eqnarray}\label{11}
\mu(r)&\mapsto&a(r)+\alpha h_{1}(r),\\\label{11a}
e^{-\varepsilon(r)}&\mapsto&e^{- b(r)}+\alpha h_{2}(r),
 \end{eqnarray}
where $h_{1}(r)$ and $h_{2}(r)$ determine the variations in temporal
and radial metric functions, respectively. It is noteworthy that the
above transformations do not alter the spherical symmetry of the
static configuration. The system of field equations
(\ref{6})-(\ref{8}) disintegrates into two independent sets under
the influence of transformed metric functions. The first set
corresponds to the original SBD system with $\alpha=0$ and encodes
the effects of the isotropic source only as
\begin{eqnarray}\nonumber
\rho&=&\frac{1}{2r^2\Upsilon(r)}\left\{e^{- b(r)}\left(r^2e^{
b(r)}V(\Upsilon)\Upsilon(r)+r^2
(-\omega_{BD})\Upsilon'^2(r)+\left(\left(r
b'(r)\right.\right.\right.\right.\\\label{12}
&-&\left.\left.\left.\left.4\right)\Upsilon'(r)
-2r\Upsilon''(r)\right)r\Upsilon(r)+2\Upsilon^2(r)\left(r b'(r)+e^{
b(r)}-1\right)\right)\right\} ,\\\nonumber
p&=&\frac{1}{2}\left\{\frac{1}{r^2\Upsilon(r)}\left(e^{-
b(r)}\left(-r^2\omega_{BD}
\Upsilon'^2(r)+\Upsilon^2(r)\left(2ra'(r)-2e^{
b(r)}+2\right)\right.\right.\right.\\\label{13}
&+&\left.\left.\left.r\Upsilon(r)
\left(ra'(r)+4\right)\Upsilon'(r)\right)\right)-V(\Upsilon)\right\},\\\nonumber
p&=&\frac{1}{4r\Upsilon(r)}\left\{e^{-
b(r)}\left(2\Upsilon(r)\left(\Upsilon'(r)\left(ra'(r)-r
 b'(r)+2\right)+2r\Upsilon''(r)\right)+\Upsilon^2(r)\right.\right.\\\nonumber
&\times&\left.\left.\left(2ra''(r)+a'(r) \left(2-r
b'(r)\right)+ra'^2(r)-2 b'(r)\right)-2re^{ b(r)}\Upsilon(r)
V(\Upsilon)\right.\right.\\\label{14}
&+&\left.\left.2r\omega_{BD}\Upsilon'^2(r)\right)\right\}.
\end{eqnarray}
The isotropic fluid distribution obeys the following conservation
equation in ($a, b$)-frame
\begin{equation}\label{14a}
T^{1'(\text{eff})}_{1}-\frac{a'(r)}{2}
(T^{0(\text{eff})}_{0}-T^{1(\text{eff})}_{1})=0.
\end{equation}
The above system is solved by assuming a viable solution (i.e.
metric potentials $a$ and $b$) corresponding to the seed source.

The second set containing effects of the additional source gives
\begin{eqnarray}\nonumber
\Theta_0^0&=&\frac{-1}{2r^2\Upsilon (r)}\left\{\left
(r\Upsilon(r)h_{2}'(r)\left(r\Upsilon'(r)+2\Upsilon
(r)\right)+h_{2}(r)\left(r^2\omega_{BD}\Upsilon'^2(r)\right.\right.\right.\\\label{15}
&+&\left.\left.\left.2r\Upsilon(r)\left(r\Upsilon
''(r)+2\Upsilon'(r)\right)+2\Upsilon^2(r)\right)\right)\right\},\\\nonumber
\Theta_1^1&=&\frac{-
h_{2}(r)}{2r^2\Upsilon(r)}\left(-r^2\omega_{BD}\Upsilon'^2(r)+r
\Upsilon(r)\left(r\mu'(r)+4\right)\Upsilon'(r)+2\Upsilon^2(r)\left(r\mu'(r)\right.\right.\\\label{16}
&+&\left.\left.1\right)\right) -\frac{ e^{-
b(r)}h_{1}'(r)\left(r\Upsilon'(r)+2\Upsilon(r)\right)}{2r},\\\nonumber
\Theta_2^2&=&\frac{-
h_{2}(r)}{4r\Upsilon(r)}\left(2\Upsilon(r)\left(\left(r\mu'(r)+2\right)\Upsilon'(r)+2
r\Upsilon''(r)\right)+\Upsilon^2(r)\left(2r\mu''(r)\right.\right.\\\nonumber
&+&\left.\left.r\mu'^2(r)+2\mu'(r)\right)+2
r\omega_{BD}\Upsilon'^2(r)\right)-\frac{
h_{2}'(r)}{4r}\left(\Upsilon(r)\left(r\mu'(r)+2\right)\right.\\\nonumber
&+&\left.2r\Upsilon'(r)\right)-\frac{ e^{-
b(r)}}{4r}\left(2rh_{1}'(r)\Upsilon'(r)+\Upsilon(r)\left(2rh_{1}''(r)+\alpha
rg'^2(r)+h_{1}'(r)\right.\right.\\\label{17}
&\times&\left.\left.\left(2ra'(r)-r b'(r)+2\right)\right)\right).
\end{eqnarray}
The conservation equation of the new source
$\Theta^{\gamma}_{\delta}$ is given as
\begin{equation}\label{17a}
\Theta^{1'(\text{eff})}_{1}
-\frac{\mu'(r)}{2}(\Theta^{0(\text{eff})}_{0}-\Theta^{1(\text{eff})}_{1})
-\frac{2}{r}(\Theta^{2(\text{eff})}_{2}-\Theta^{1(\text{eff})}_{1})
=\frac{h_{1}'(r)}{2}(T^{0(\text{eff})}_{0}-T^{1(\text{eff})}_{1}),
\end{equation}
where
\begin{eqnarray}\nonumber
\Theta^{0(\text{eff})}_0&=&\frac{1}{\Upsilon}\left(\Theta^0_0+\frac{1}{2}
h_{2}'(r)\Upsilon'(r)+h_{2}(r)\Upsilon''+\frac{\omega_{BD}h_{2}(r)\Upsilon'^2}{2\Upsilon}
+\frac{2h_{2}(r)\Upsilon '(r)}{r}\right),\\\nonumber
\Theta^{1(\text{eff})}_1&=&\frac{1}{\Upsilon}\left(\Theta^1_1+\frac{1}{2r\Upsilon}
e^{- b(r)}\Upsilon'(r)\left(h_{2}(r)e^{
b(r)}\left(\Upsilon(r)\left(r\mu'(r)+4\right)-r
\omega_{BD}\right.\right.\right.\\\nonumber
&\times&\left.\left.\left.\Upsilon'(r)\right)+r\Upsilon(r)h_{1}'(r)\right)\right),\\\nonumber
\Theta^{2(\text{eff})}_2&=&\frac{1}{\Upsilon}\left(\Theta^2_2+\frac{1}{2r\Upsilon}
e^{- b(r)}\left(r\Upsilon(r)\Upsilon'(r)\left(e^{
b(r)}h_{2}'(r)+h_{1}'(r)\right)+h_{2}(r)\right.\right.\\\nonumber&\times&\left.\left.
e^{
b(r)}\left(\Upsilon(r)\left(\left(r\mu'(r)+2\right)\Upsilon'(r)+2r\Upsilon''(r)\right)+r
\omega_{BD}\Upsilon'^2(r)\right)\right)\right).
\end{eqnarray}
The effective energy-momentum tensor
$T^{\gamma(\text{eff})}_{\delta}$ is conserved in
($\mu,\varepsilon$)-coordinate system as
\begin{equation}\label{18a}
\nabla_{\gamma}T^{\gamma(\text{eff})}_{\beta}=\nabla_{\gamma}^{(a,
b)}T^{\gamma(\text{eff})}_{\beta}
-\frac{h_{1}'(r)}{2}(T^{0(\text{eff})}_{0}-T^{1(\text{eff})}_{1})\delta^1_{\beta},
\end{equation}
where $\nabla_{\gamma}^{(a, b)}$ denotes the divergence in ($a,
b$)-frame of reference. Moreover, Eqs.(\ref{1444}) and (\ref{17a})
yield
\begin{equation}\label{19}
\nabla_{\gamma}^{(a, b)}T^{\gamma(\text{eff})}_{\beta}=0,\quad
\nabla_{\gamma}\Theta^{\gamma(\text{eff})}_{\beta}
=\frac{h_{1}'(r)}{2}(T^{0(\text{eff})}_{0}-T^{1(\text{eff})}_{1})\delta^1_{\beta}.
\end{equation}

It can be directly deduced from Eqs.(\ref{18a}) and (\ref{19}) that
the sources $T^{(m)}_{\gamma\delta}$ and $\Theta_{\gamma\delta}$ are
not conserved allowing the exchange of energy between them. However,
the total energy and momentum of the setup remain unaltered. Thus,
the EGD technique can be used to decouple these sources as long as
energy transfer from one setup to the other is possible. This
feature differentiates the EGD approach from MGD technique where a
radial transformation decouples the two individually conserved
sources. However, in the case of vacuum or barotropic fluid, EGD can
successfully decouple the matter sources interacting only
gravitationally. Combining the two systems (\ref{12})-(\ref{14}) and
(\ref{15})-(\ref{17}) yields the state determinants of the extended
solution as $\rho+\alpha\Theta^0_0,~p_r=p-\alpha\Theta^1_1$ and
$p_\perp=p-\alpha\Theta^2_2$. Shifting the components of
$\Theta^\gamma_\delta$ to the other side, the density and pressure
of the extended solution are completely determined as
\begin{eqnarray}\nonumber
\rho&=&\frac{e^{- b(r)}}{2r^2\Upsilon
(r)}\left(-r\Upsilon(r)\left(\Upsilon'(r)\left(\alpha r e^{
b(r)}h_{2}'(r)+4\alpha h_{2}(r)e^{ b(r)}-r
b'(r)+4\right)\right.\right.\\\nonumber&+&\left.\left.2r\Upsilon
''(r)\left(\alpha h_{2}(r)e^{ b(r)}+1\right)\right)-2\Upsilon^2(r)
\left(\alpha re^{ b(r)}h_{2}'(r)+\alpha h_{2}(r)e^{
b(r)}\right.\right.\\\nonumber&-&\left.\left.r b'(r)-e^{ b
(r)}+1\right)+r^2(-\omega_{BD})\Upsilon'^2(r)\left(\alpha
h_{2}(r)e^{ b (r)}+1\right)\right.\\\label{28}&+&\left.r^2e^{
b(r)}\Upsilon(r)V(\Upsilon)\right),\\\nonumber
p_r&=&\frac{\Upsilon(r)}{r^2}\left(\left(\alpha h_{2}(r)+e^{- b
(r)}\right)\left(\alpha
rh_{1}'(r)+ra'(r)+1\right)-1\right)-\frac{1}{2r\Upsilon
(r)}\\\nonumber&\times&\left(\Upsilon'(r)\left(\alpha h_{2}(r)+e^{-
b(r)}\right)\left(r\omega_{BD}\Upsilon'(r)-\Upsilon(r) \left(\alpha
rh_{1}'(r)+ra'(r)+4\right)\right)\right)\\\label{29}&-&\frac{V(\Upsilon)}{2},\\\nonumber
p_\perp&=&\left(\alpha h_{2}(r)+e^{-
b(r)}\right)\left(\frac{1}{2}\Upsilon'(r)\left(\frac{\alpha e^{
b(r)}h_{2}'(r)- b'(r)}{\alpha h_{2}(r)e^{ b(r)}+1}+\alpha
h_{1}'(r)+a'(r)\right.\right.\\\nonumber&+&\left.\left.\frac{2}{r}\right)+\Upsilon''(r)+\frac{\omega_{BD}\Upsilon'^2(r)}{2
\Upsilon(r)}\right)+\frac{1}{2}\Upsilon(r)\left(\alpha h_{2}(r)+e^{-
b (r)}\right)\left((\left(\alpha e^{
b(r)}h_{2}'(r)\right.\right.\\\nonumber&-&\left.\left. b'(r)\right)
\left(\alpha h_{1}'(r)+a'(r)\right))(2\alpha h_{2}(r)e^{
b(r)}+2)^{-1}+\frac{1}{r}(\frac{\alpha e^{ b(r)}h_{2}'(r)-
b'(r)}{\alpha h_{2}(r)e^{ b(r)}+1}\right.\\\label{30}&+&\left.\alpha
h_{1}'(r)+a'(r))+\alpha g''+\frac{1}{2} \left(\alpha
h_{1}'(r)+a'^2(r)\right)+a''(r)\right)-\frac{V(\Upsilon)}{2}.
\end{eqnarray}

\section{Extended Schwarzschild Solutions}

The process of extracting solutions of the field equations is
simplified by splitting the original system into two sets. The first
set governed by the isotropic source is completely specified by
assuming a suitable metric leaving fewer unknowns in the second set.
For this purpose, we consider the Schwarzschild metric expressed as
\begin{equation}\label{20a}
ds^2=(1-\frac{2M}{r})dt^2-\frac{1}{(1-\frac{2M}{r})}dr^2
-r^2(d\theta^2+\sin^2\theta d\phi^2),
\end{equation}
with a singularity at $r=0$ hidden behind an event horizon at
$r=2M$. Consequently, $e^a=e^{- b}=1-\frac{2M}{r}$. According to the
no-hair theorem, information about the BH is lost behind the event
horizon as physical state of matter is unknown beyond this boundary.
However, efforts have been made to study BHs in different
perspectives in order to avoid this theorem \cite{6}. The addition
of scalar field \cite{44} or another generic source of matter
\cite{26} in the vacuum leads to different BH solutions known as
hairy BHs (with mass $M$ and a discrete set of charges as primary
hair). The major benefit of the EGD scheme is the transformation in
temporal as well as radial metric components which increases the
probability of hairy BH solutions with different horizons.

As there are five unknowns
$(h_{1}(r),~h_{2}(r),~\Theta^0_0,~\Theta^1_1,~\Theta^2_2)$ in the
system (\ref{15})-(\ref{17}), we apply two constraints to obtain the
extended solution. In order to have a  well-defined causal structure
of the resultant spacetime, it is necessary that the causal horizon
($e^{-\varepsilon}=0$) either covers the Killing horizon ($e^\mu=0$)
or coincides with it. Therefore, the first constraint is applied to
the metric potentials as
\begin{equation}\label{40}
\mu=-\varepsilon.
\end{equation}
which leads to coinciding Killing and causal horizons. Setting
$e^{-\varepsilon}=0$ implies that both Killing and causal horizons
occur at $r=2M$. The second constraint is applied to
$\Theta$-components through a linear EoS expressed as
\begin{equation}\label{42}
\Theta_0^0=a_1\Theta_1^1+a_2\Theta_2^2,
\end{equation}
where $a_1$ and $a_2$ are constants. This EoS has been considered
previously to evaluate extended solutions through the decoupling
method \cite{14}. Applying the condition (\ref{40}) to
Eqs.(\ref{11}) and (\ref{11a}) yields the following relation between
the deformation functions
\begin{equation}\label{41}
h_{2}(r)=-\frac{(2M-r)\left(e^{\alpha h_{1}(r)}-1\right)}{\alpha r}.
\end{equation}

The presence of an essential singularity is confirmed through
Kretschmann scalar which is evaluated as
\begin{eqnarray}\nonumber
\mathcal{K}&=&\frac{1}{r^6}\left[\alpha^2r^4(r-2M)^2e^{2\alpha
h_1(r)}h_1''^2+8\alpha Mr^2(2M-r)e^{2\alpha
h_1(r)}h_1''(r)+\alpha^4r^4(r\right.\\\nonumber
&-&\left.2M)^2e^{2\alpha h_1(r)}
h_1'^4-8\alpha^3Mr^3(2M-r)e^{2\alpha
h_1(r)}h_1'^3+2\alpha^2r^2e^{2\alpha h_1(r)}h_1'^2\left(\alpha
r^2\right.\right.\\\nonumber &\times&\left.\left.(r-2M)^2
h_1''(r)+2\left(12M^2-6Mr+r^2\right)\right)-8\alpha Mre^{2\alpha
h_1(r)} h_1'(r)\left(\alpha r^2\right.\right.\\\nonumber
&\times&\left.\left.(2M-r)h_1''(r)+8M-2r\right)+4\left(\left(12M^2-4
Mr+r^2\right)e^{2\alpha h_1(r)}\right.\right.\\\nonumber
&+&\left.\left.2r(2M-r)e^{\alpha h_1(r)}+r^2\right)\right].
\end{eqnarray}
In the subsequent subsections, $\mathcal{K}$ is plotted to indicate
the presence of a singularity at $r=0$. We compute decoupled
solutions by choosing different values of the constants $a_1$ and
$a_2$ corresponding to different scenarios. Moreover, we set
$V(\Upsilon)=\frac{1}{2}m_{\Upsilon}^2\Upsilon^2$, where
$m_{\Upsilon}$ is the scalar field mass. The restriction on the
values of the coupling parameter due to weak-field tests are waived
off for $m_\Upsilon>10^{-4}$ (in dimensionless units). Thus, we
determine the massive scalar field by solving the wave equation
numerically for $m_\Upsilon=0.1$ and $\omega_{BD}=60$. The behavior
of the obtained BH solutions is checked for
$\alpha=-0.4,~-0.5,~-0.7$. It is noteworthy to mention here that
under the applied constraints, the positive behavior of density is
achieved for negative values of the decoupling parameter only.

\subsection{Case I: Traceless $\Theta_\delta^\gamma$}

\begin{figure}\center
\epsfig{file=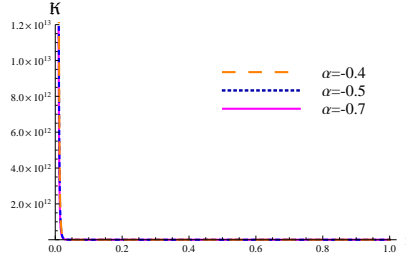,width=0.4\linewidth} \caption{Plot of
$\mathcal{K}$ for case I.}\label{401}
\end{figure}

\begin{figure}\center
\epsfig{file=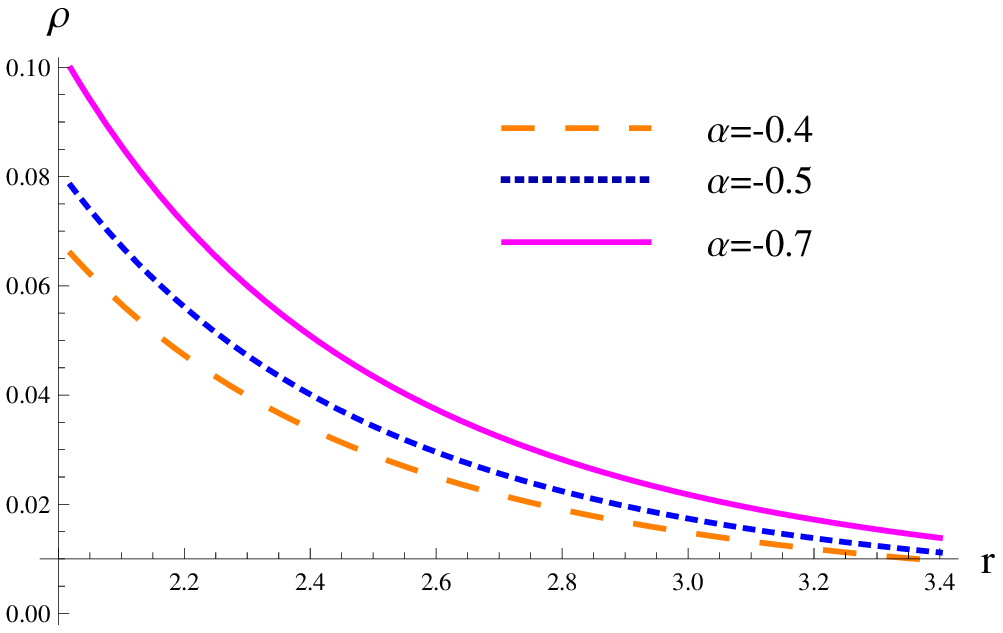,width=0.4\linewidth}\epsfig{file=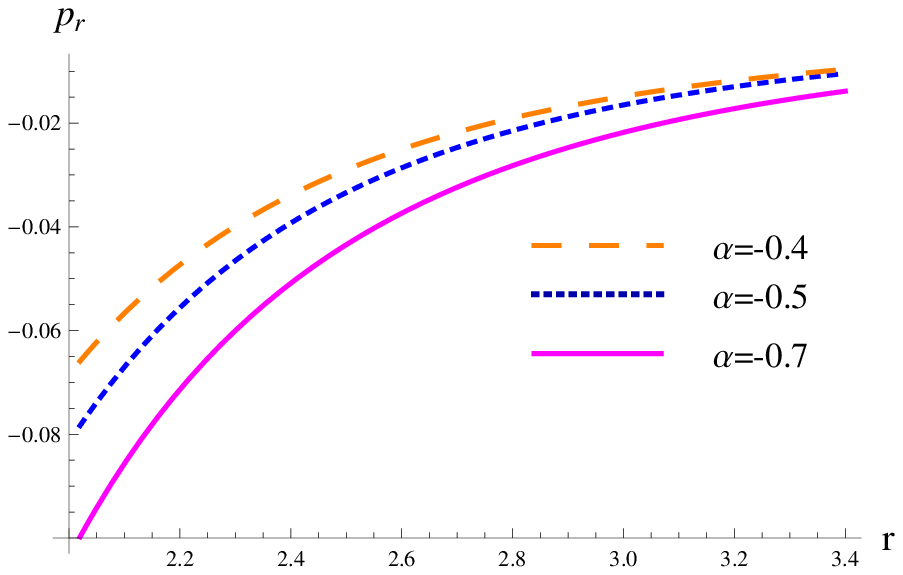,width=0.4\linewidth}
\epsfig{file=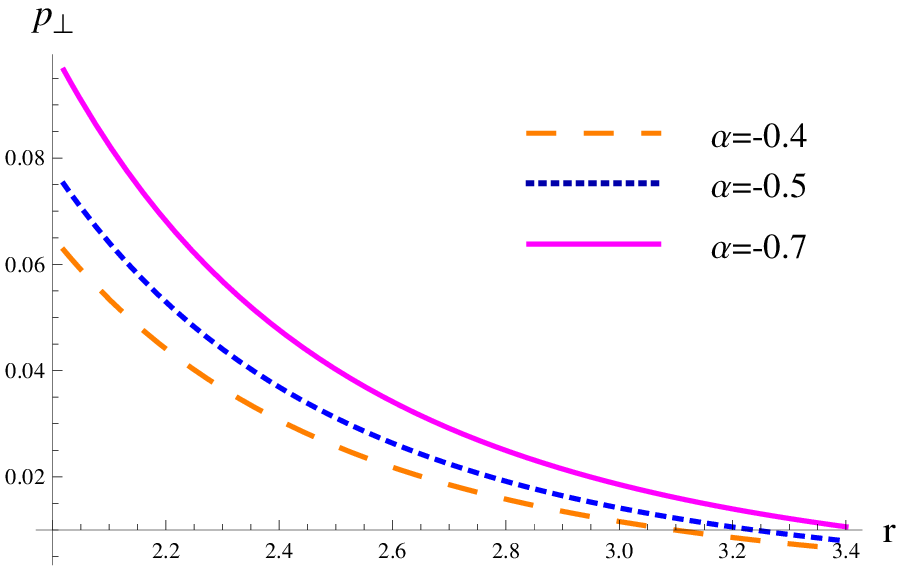,width=0.4\linewidth} \caption{Plots of matter
variables for case I.}\label{402}
\end{figure}

\begin{figure}\center
\epsfig{file=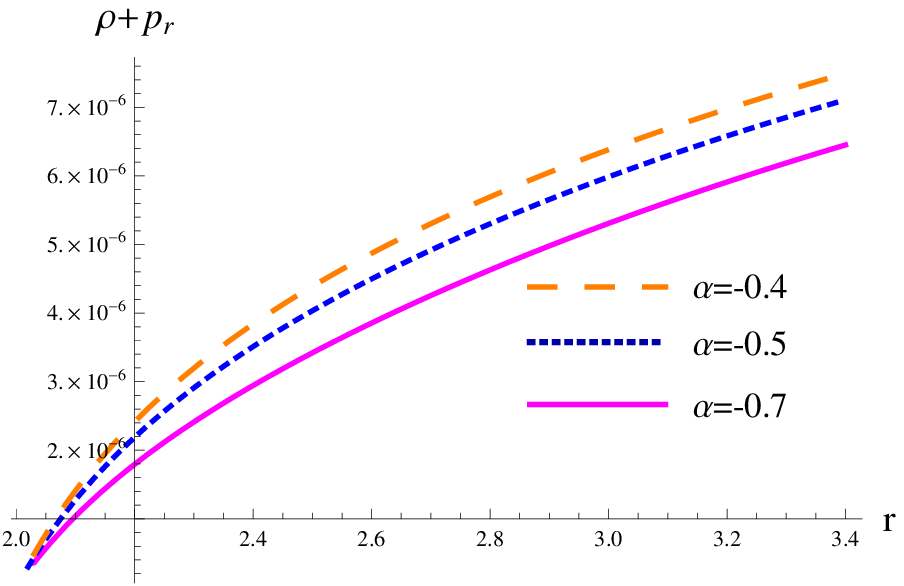,width=0.4\linewidth}\epsfig{file=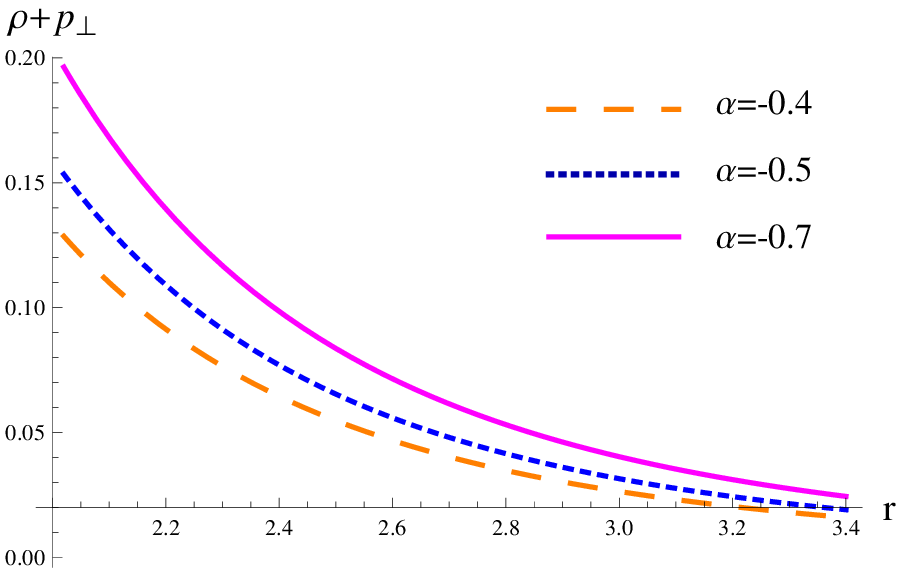,width=0.4\linewidth}
\epsfig{file=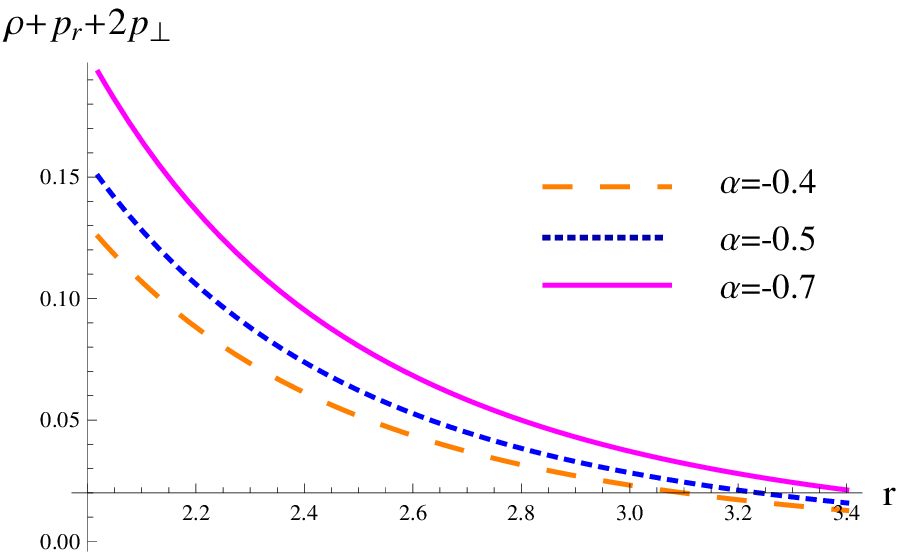,width=0.4\linewidth}\epsfig{file=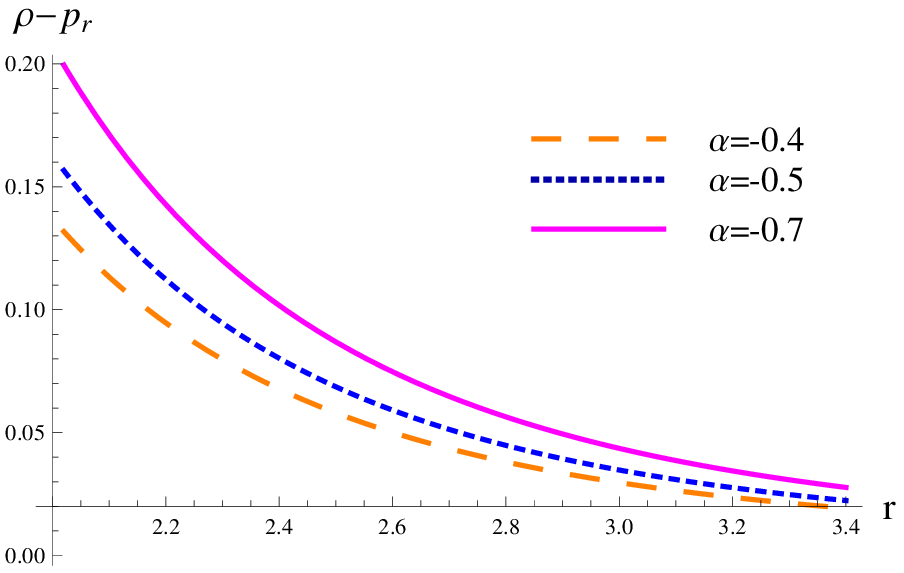,width=0.4\linewidth}
\epsfig{file=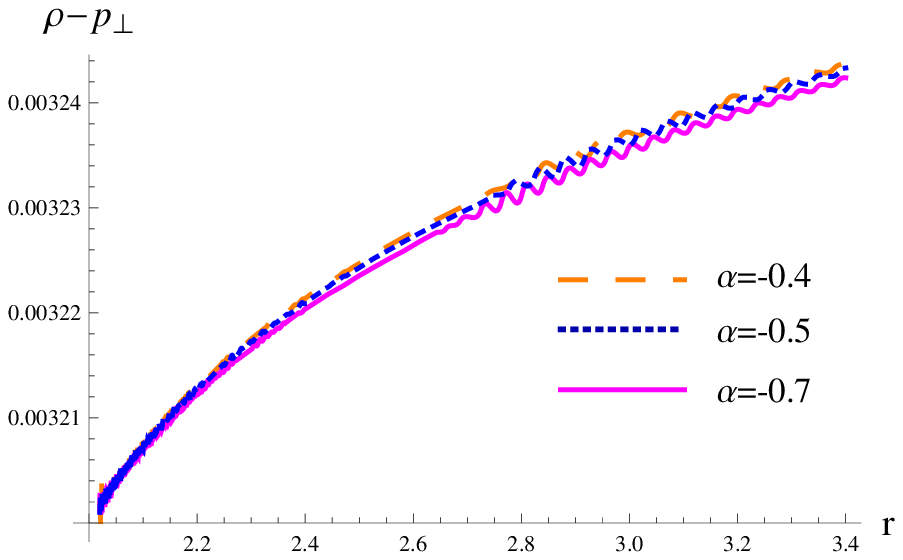,width=0.4\linewidth} \caption{Energy conditions
for case I.}\label{403}
\end{figure}

\begin{figure}\center
\epsfig{file=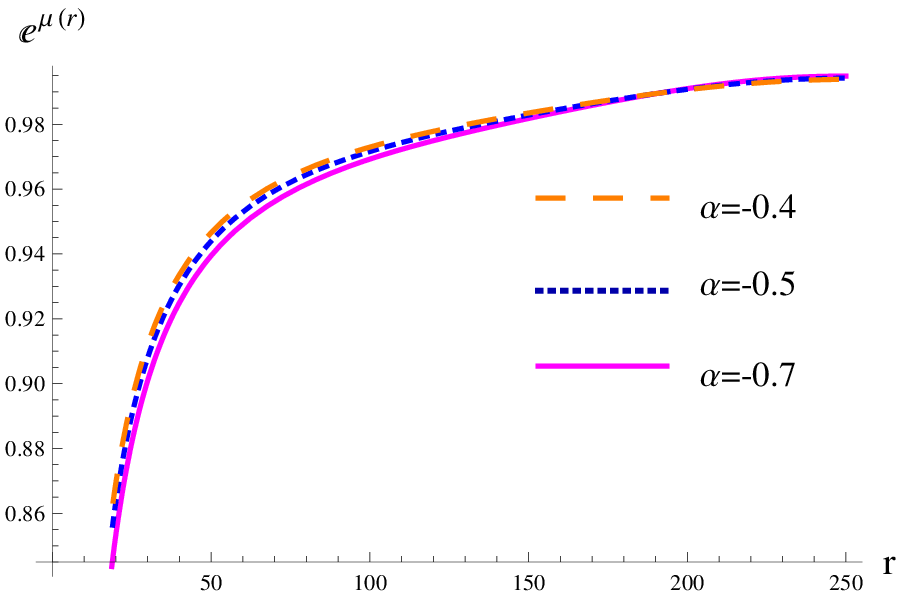,width=0.4\linewidth}\epsfig{file=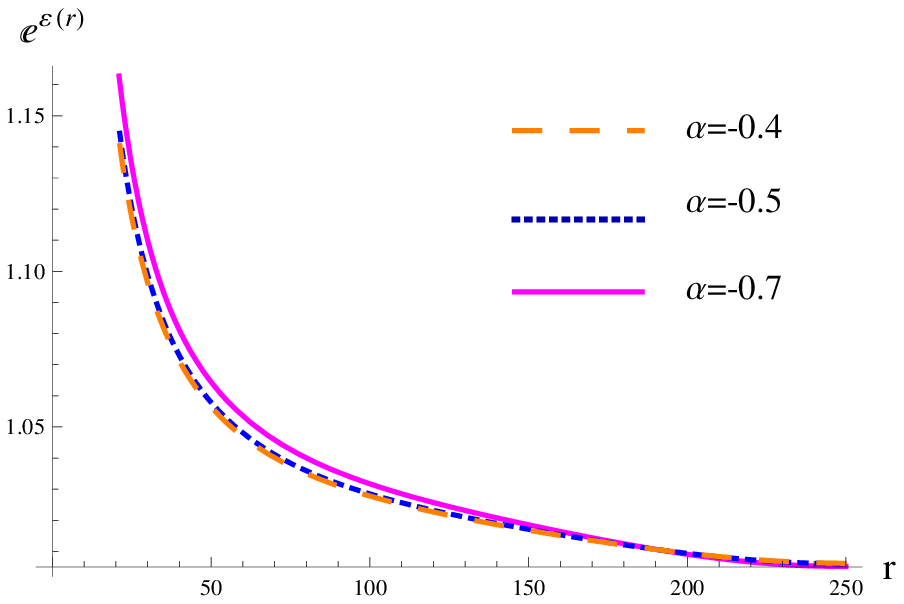,width=0.4\linewidth}
\caption{Metric potentials for case I.}\label{404}
\end{figure}

The additional source has a traceless energy-momentum tensor when
\begin{equation}\nonumber
\Theta_0^0+\Theta_1^1=-2\Theta_2^2,
\end{equation}
(since $\Theta_2^2=\Theta_3^3$), i.e., $\Theta_\delta^\gamma$ is
traceless when $a_1=-1$ and $a_2=-2$ in Eq.(\ref{42}) which yields
\begin{eqnarray}\nonumber
&&\frac{1}{r\Upsilon(r)}\left(e^{- b(r)}\left(\alpha\Upsilon(r)r
\left(3\Upsilon'(r)r \left(e^{
b(r)}h_{2}'(r)+h_{1}'(r)\right)+\Upsilon(r)\left(e^{
b(r)}h_{2}'(r)\right.\right.\right.\right.\\\nonumber
&&\left.\left.\left.\left.\times\left(\alpha r
h_{1}'(r)+ra'(r)+4\right)+2rh_{1}''(r)+h_{1}'(r)\left(\alpha r
h_{1}'(r)+2ra'(r)-r \right.\right.\right.\right.\right.\\\nonumber
&&\left.\left.\left.\left.\left.\times
b'(r)\right)+4\right)\right)+\alpha h_{2}(r)e^{ b
(r)}\left(3r\Upsilon(r)\left(\Upsilon'(r)\left(\alpha
rh_{1}'(r)+ra'(r)+4\right)\right.\right.\right.\right.\\\nonumber
&&\left.\left.\left.\left.+2r\Upsilon''(r)\right)+\Upsilon^2(r)\left(r\left(r
\left(2\left(\alpha h_{1}''(r)+a''(r)\right)+a'^2(r)\right)+\alpha^2
rg'^2(r)\right.\right.\right.\right.\right.\\\label{43}
&&\left.\left.\left.\left.\left.+2\alpha
h_{1}'(r)\left(ra'(r)+2\right)+4a'(r)\right)+4\right)+2r^2\omega_{BD}
\Upsilon'^2(r)\right)\right)\right)=0.
\end{eqnarray}
The deformation function $h_{1}(r)$ is evaluated by solving the
above equation numerically along with the wave equation subject to
the initial conditions
$\Upsilon(2M)=0.8,~\Upsilon'(2M)=0.1,~g(2M)=1$ and $g'(2M)=0.1$. The
Kretschmann scalar shown in Figure \ref{401} approaches to infinity
when $r\rightarrow0$. Thus, a singularity exists at $r=0$. The
graphical analysis of state variables is done in the region
accessible to an outer-observer with $M=1$ (Figure \ref{402}). The
energy density and tangential pressure increase as the decoupling
parameter decreases. However, radial pressure is directly
proportional to $\alpha$. It is noted that positive energy density
is obtained when radial pressure is negative. The matter source is
normal if the state determinants adhere to four energy bounds. The
energy conditions (null, weak, strong and dominant) in the framework
of SBD theory are, respectively, expressed as \cite{45}
\begin{eqnarray}\nonumber
&&\text{NEC:}\quad\rho+p_r\geq0,\quad\rho+p_\perp\geq0,\\
&&\text{WEC:}\quad\rho\geq0,\quad\rho+p_r\geq0,\quad\rho+p_\perp\geq0,\\
&&\text{SEC:}\quad\rho+p_r+2p_\perp\geq0,\\
&&\text{DEC:}\quad\rho-p_r\geq0,\quad \rho-p_\perp\geq0.
\end{eqnarray}
When the above conditions are violated the matter is termed as
exotic. Figure \ref{403} shows that all bounds on energy density and
pressure components are satisfied in the considered setup. The plot
of metric potentials in Figure \ref{404} indicates that the new
solution preserves asymptotic flatness for large values of the
radial coordinate.
\begin{figure}\center
\epsfig{file=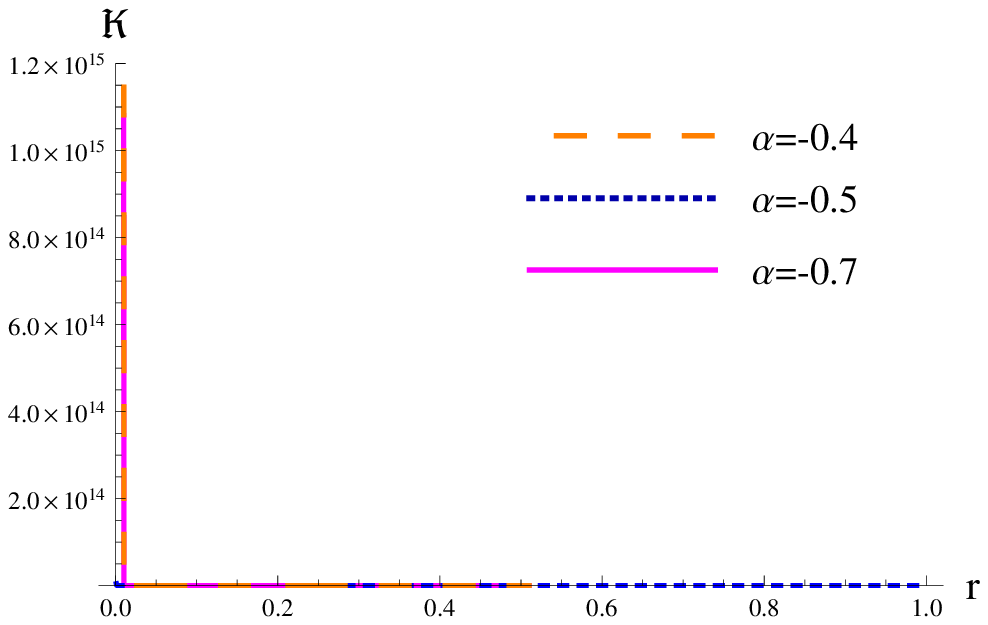,width=0.4\linewidth} \caption{Plot of
$\mathcal{K}$ for case II.}\label{405}
\end{figure}

\subsection{Case II: Barotropic Equation of State}
\begin{figure}\center
\epsfig{file=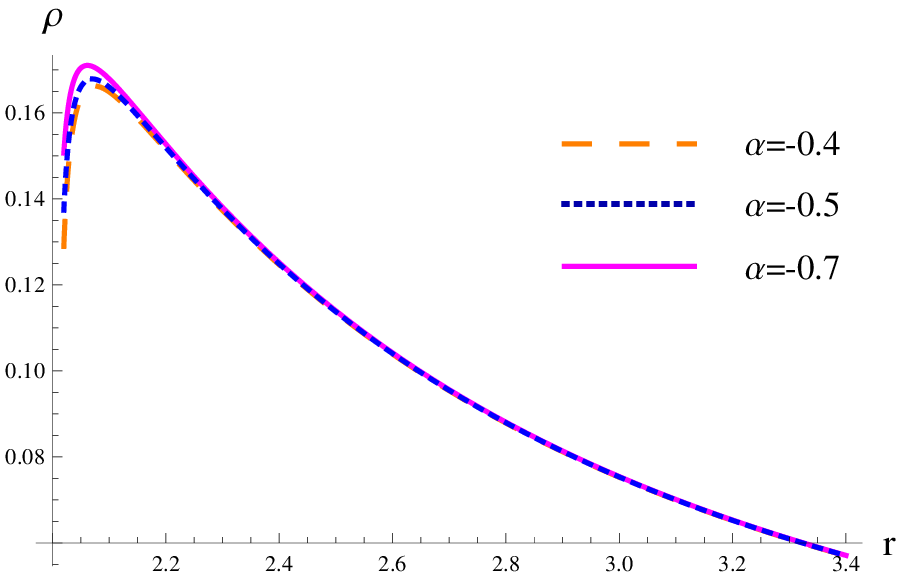,width=0.4\linewidth}\epsfig{file=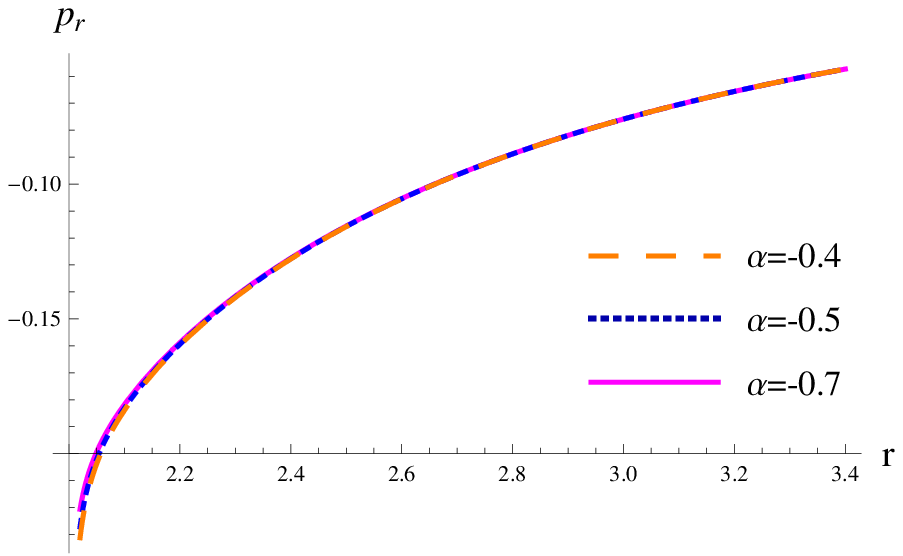,width=0.4\linewidth}
\epsfig{file=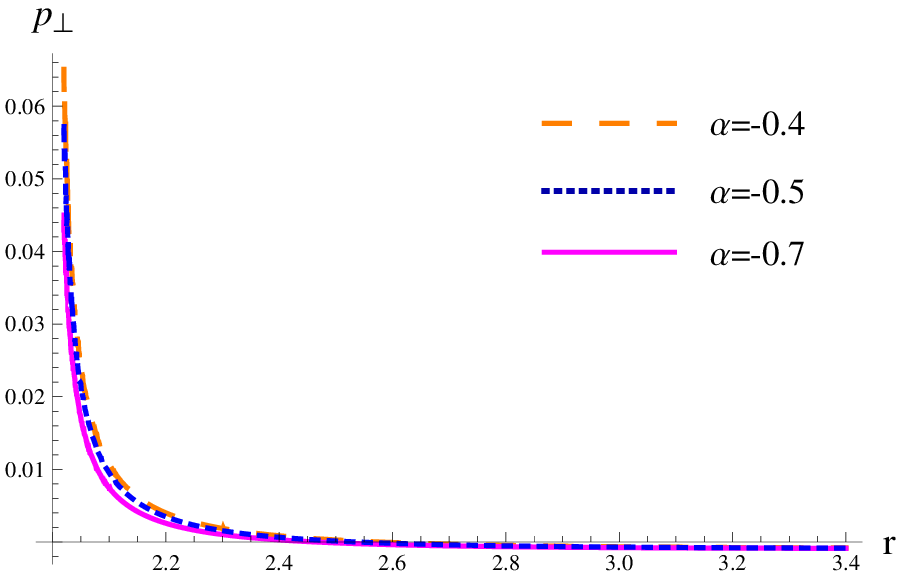,width=0.4\linewidth} \caption{Matter variables
for case II.}\label{406}
\end{figure}

The new source is polytropic if it satisfies the EoS
\begin{equation}\nonumber
\Theta_1^1=K(\Theta_0^0)^\Gamma,
\end{equation}
where $\Gamma=1+\frac{1}{n}$ is the polytropic index and $K$ is a
positive parameter containing information about temperature.
Different values of the polytropic index correspond to different
types of fluids. We proceed by taking the simplest case of
barotropic fluid (isothermal sphere of gas) associated with
$\Gamma=1$. The resulting EoS is equivalent to Eq.(\ref{42}) for
$a_1=-\frac{1}{K}$ and $a_2=0$ which is expressed as
\begin{eqnarray}\nonumber
&&\frac{1}{Kr\Upsilon(r)}\left(\alpha e^{-
b(r)}\left(r\Upsilon(r)\left(r\Upsilon'(r)+2\Upsilon(r)\right)\left(K
e^{ b(r)}h_{2}'(r)+h_{1}'(r)\right)\right.\right.\\\nonumber
&&\left.\left.+h_{2}(r)e^{ b(r)}\left(r\Upsilon(r)
\left(\Upsilon'(r)\left(\alpha
rh_{1}'(r)+4K+ra'(r)+4\right)+2Kr\Upsilon''(r)\right)
\right.\right.\right.\\\nonumber
&&\left.\left.\left.\left.+2\Upsilon^2(r)\left(\alpha\right.
rh_{1}'(r)+K+ra'(r)+1\right)+(K-1)r^2\omega_{BD}\Upsilon'^2(r)\right)\right)\right)=0.\\\label{44}
\end{eqnarray}
Employing Eq.(\ref{41}) and the initial conditions used in case I,
Eqs.(\ref{2}) and (\ref{44}) are solved simultaneously for
$\Upsilon(r)$ and $h_{1}(r)$, respectively with $M=1$.

\begin{figure}\center
\epsfig{file=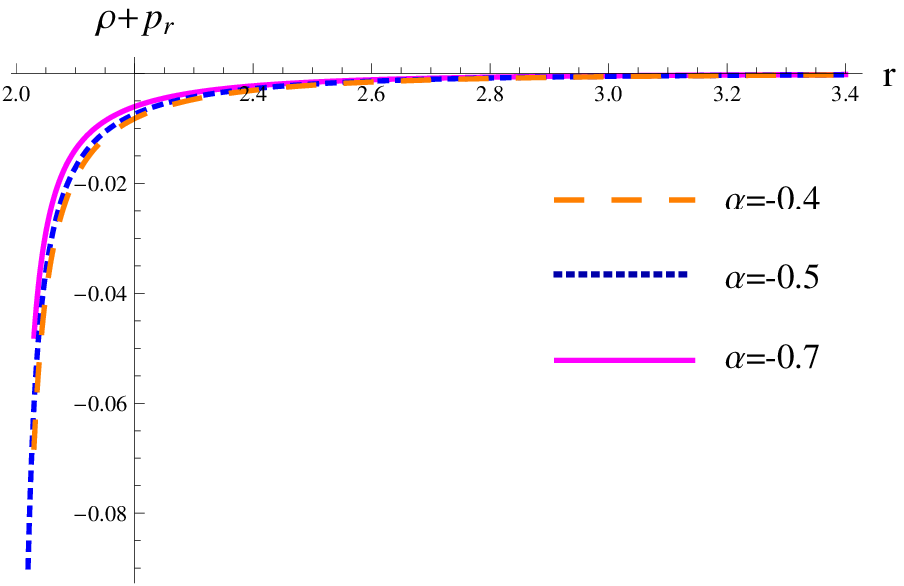,width=0.35\linewidth}\epsfig{file=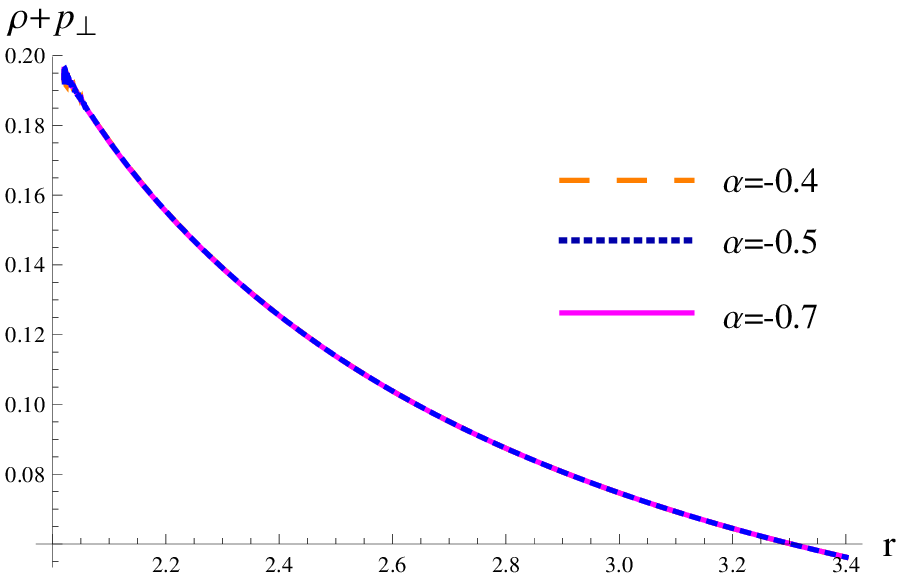,width=0.4\linewidth}
\epsfig{file=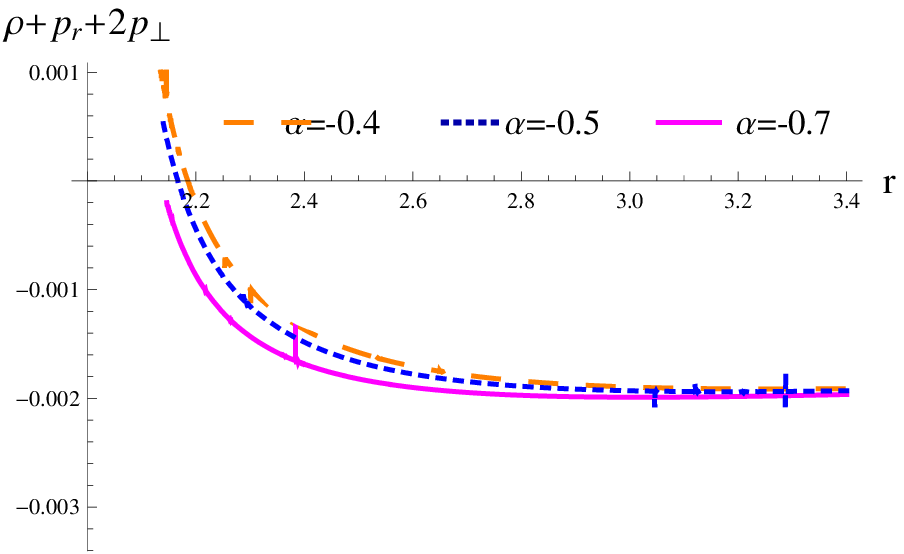,width=0.4\linewidth}\epsfig{file=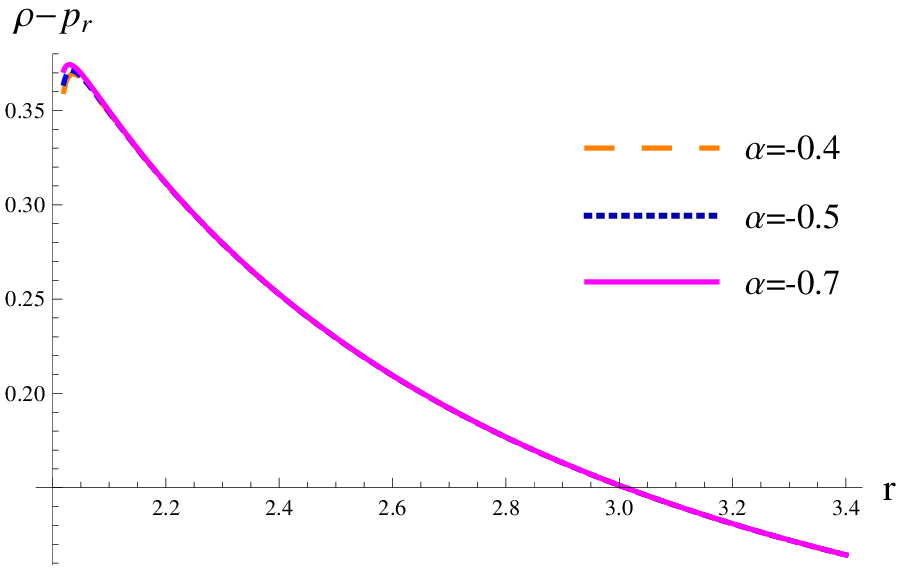,width=0.4\linewidth}
\epsfig{file=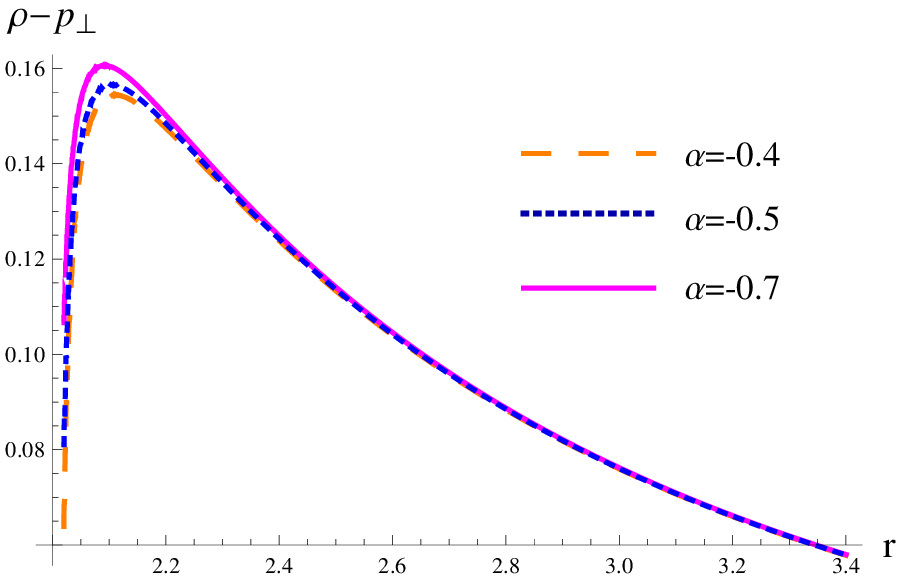,width=0.35\linewidth} \caption{Energy
conditions for case II.}\label{407}
\end{figure}

\begin{figure}\center
\epsfig{file=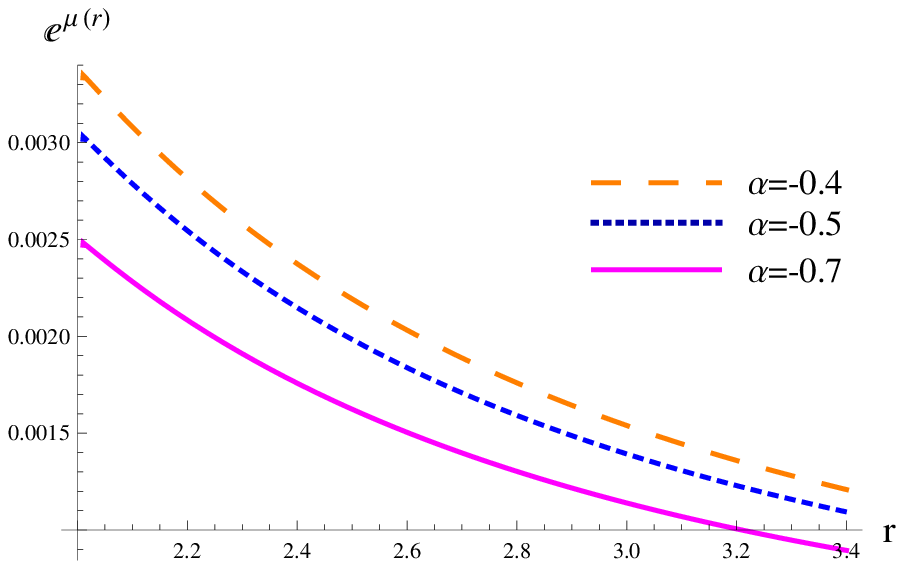,width=0.35\linewidth}\epsfig{file=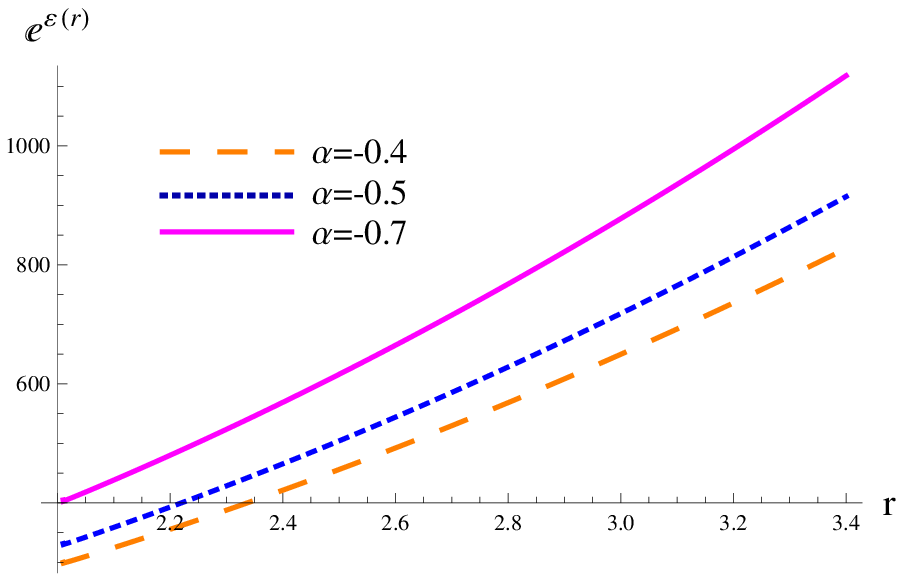,width=0.35\linewidth}
\caption{Metric potentials for case II.}\label{408}
\end{figure}

Figure \ref{405} demonstrates the presence of a singularity at $r=0$
in the current setup. The plots of energy density and pressure
components are displayed in Figure \ref{406} for $K=0.01$. The
celestial object becomes less dense for higher values of $\alpha$.
The density increases to a maximum and then decreases monotonically
for $r>2M$. Moreover, the radial pressure decreases while tangential
pressure increases as the decoupling parameter takes on higher
values. The plots in Figure \ref{407} indicate that the extended
solution fails to satisfy the energy conditions as $\rho+p_r<0$ and
$\rho+p_r+2p_\perp<0$ for the chosen values of $\alpha$. Thus, the
unknown source $\Theta^\gamma_\delta$ can be treated as exotic
matter in this case. Finally, the metric potentials representing the
spacetime of this setup are shown in Figure \ref{408}. It can be
clearly observed that trends of $e^\mu$ and $e^\varepsilon$ do not
approach 1 and thus, disobey the criterion of asymptotic flatness.

\subsection{Case III: A Particular Solution}

\begin{figure}\center
\epsfig{file=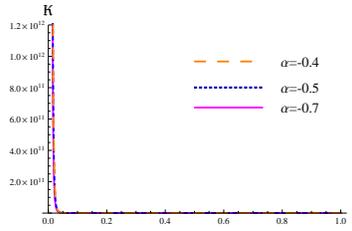,width=0.35\linewidth} \caption{Plot of
$\mathcal{K}$ for case III.}\label{409}
\end{figure}

\begin{figure}\center
\epsfig{file=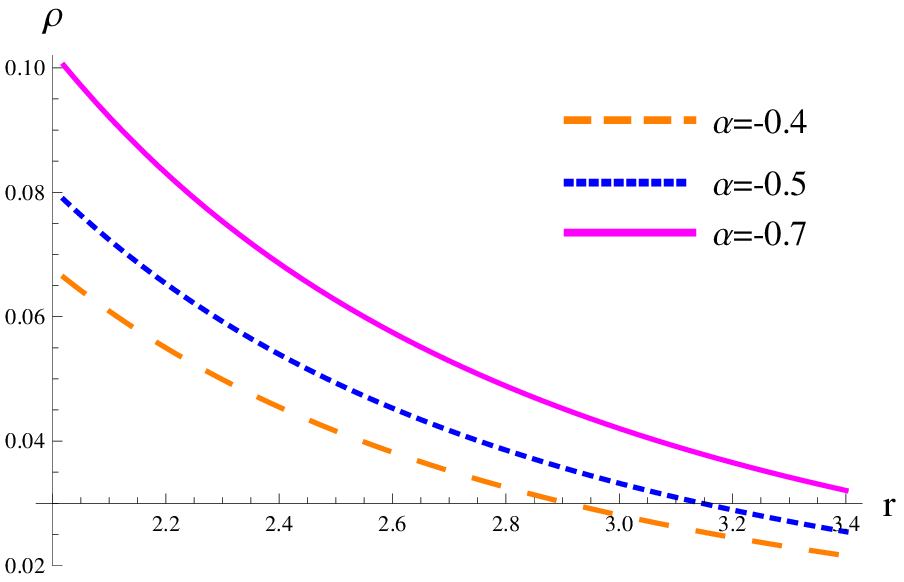,width=0.4\linewidth}\epsfig{file=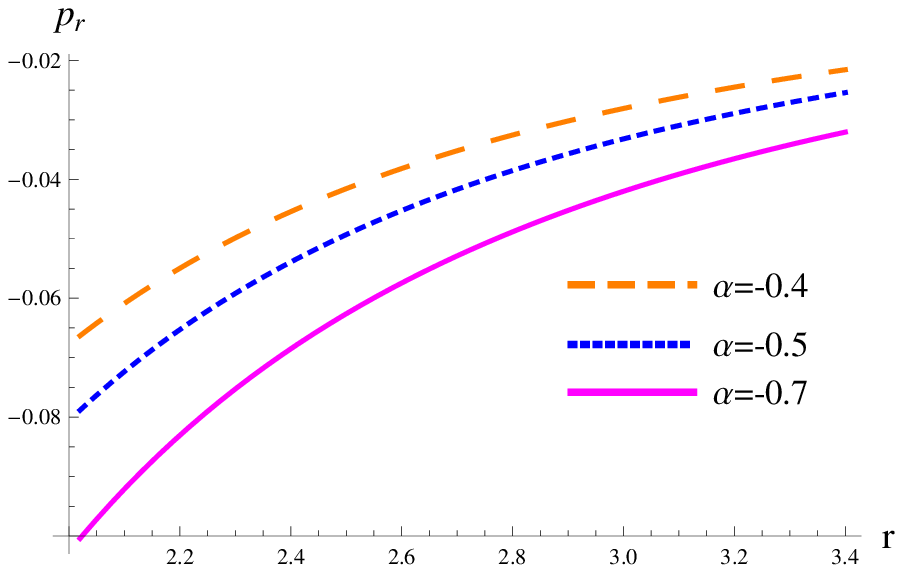,width=0.4\linewidth}
\epsfig{file=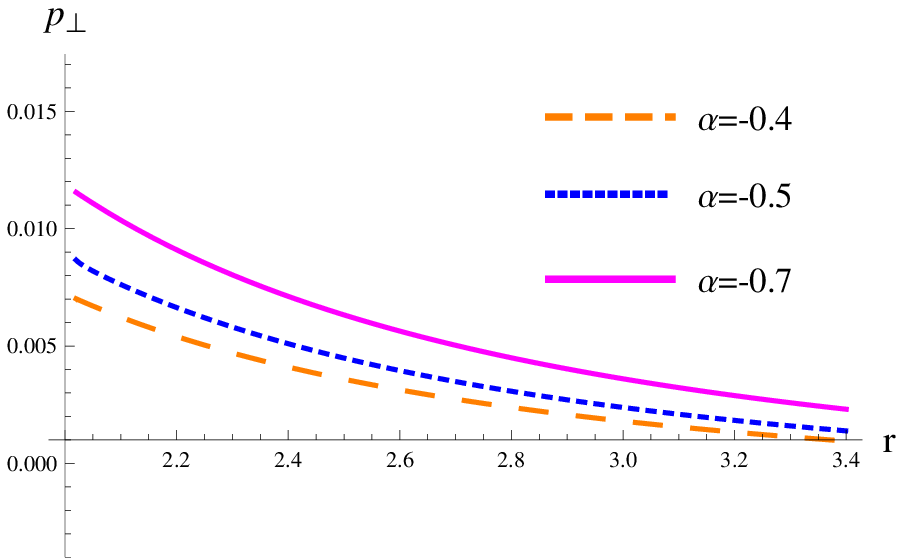,width=0.4\linewidth} \caption{Matter variables
for case III.}\label{410}
\end{figure}

\begin{figure}\center
\epsfig{file=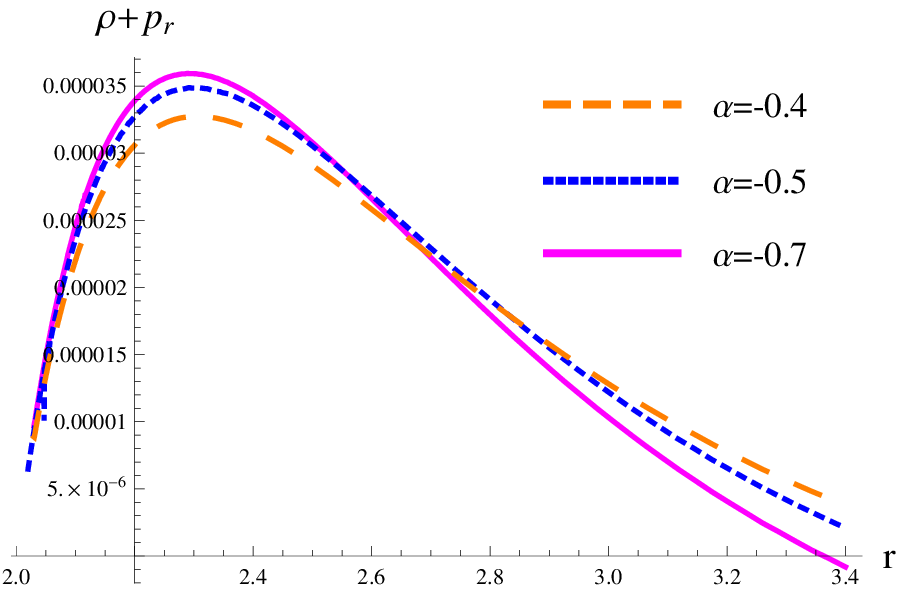,width=0.4\linewidth}\epsfig{file=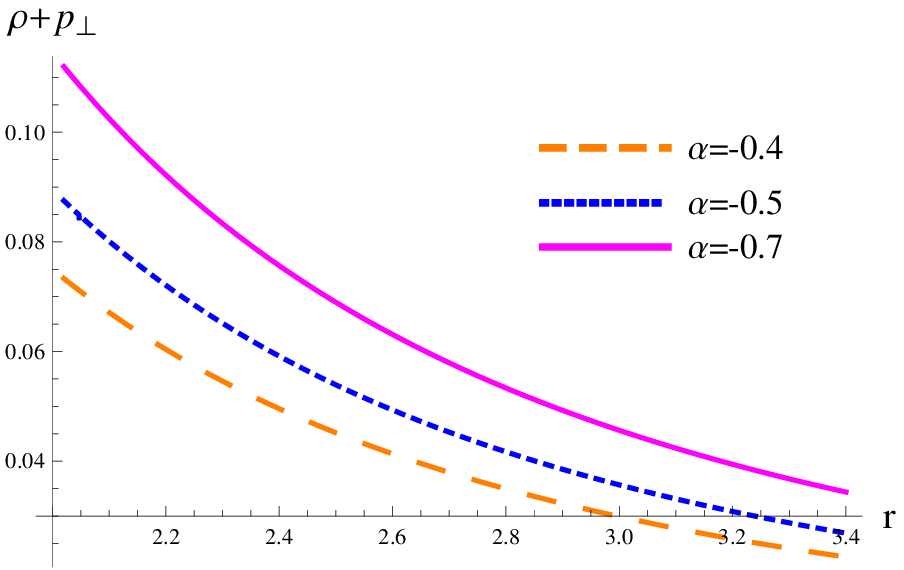,width=0.35\linewidth}
\epsfig{file=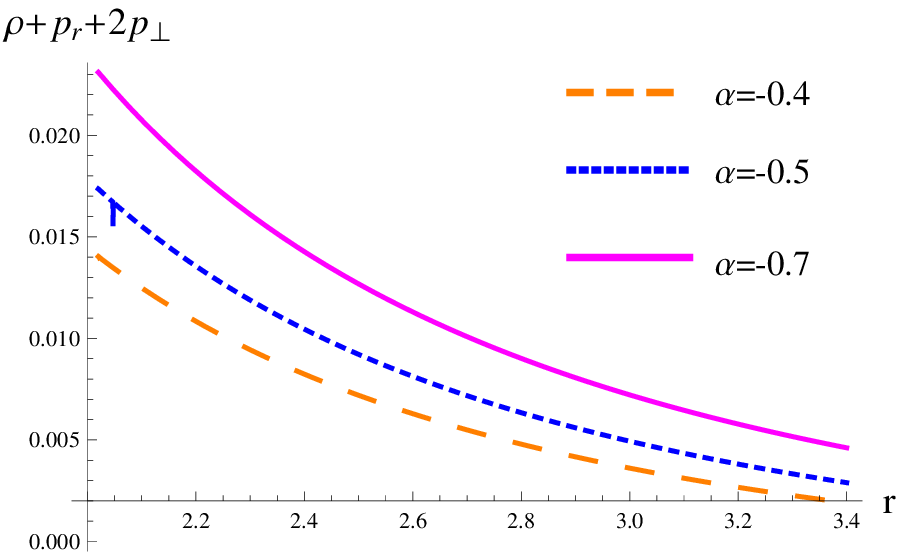,width=0.35\linewidth}\epsfig{file=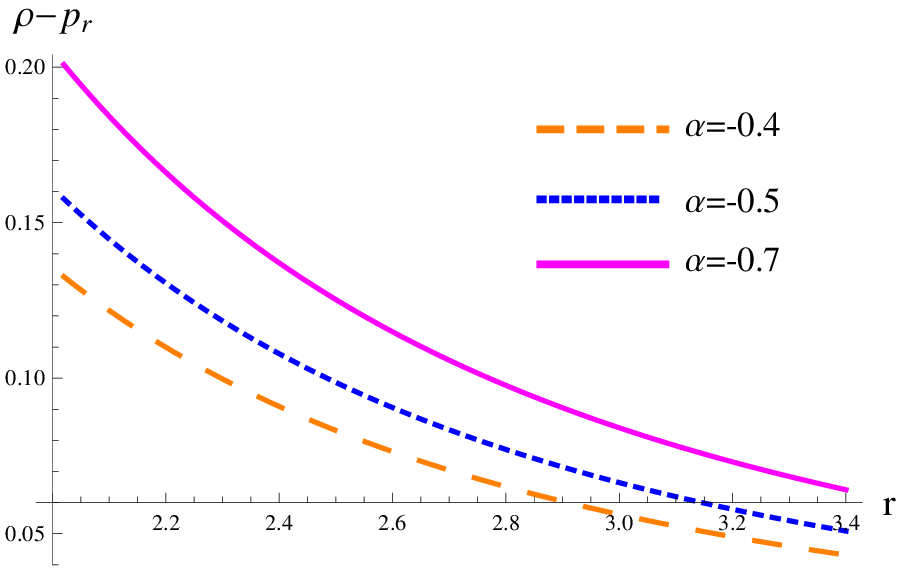,width=0.35\linewidth}
\epsfig{file=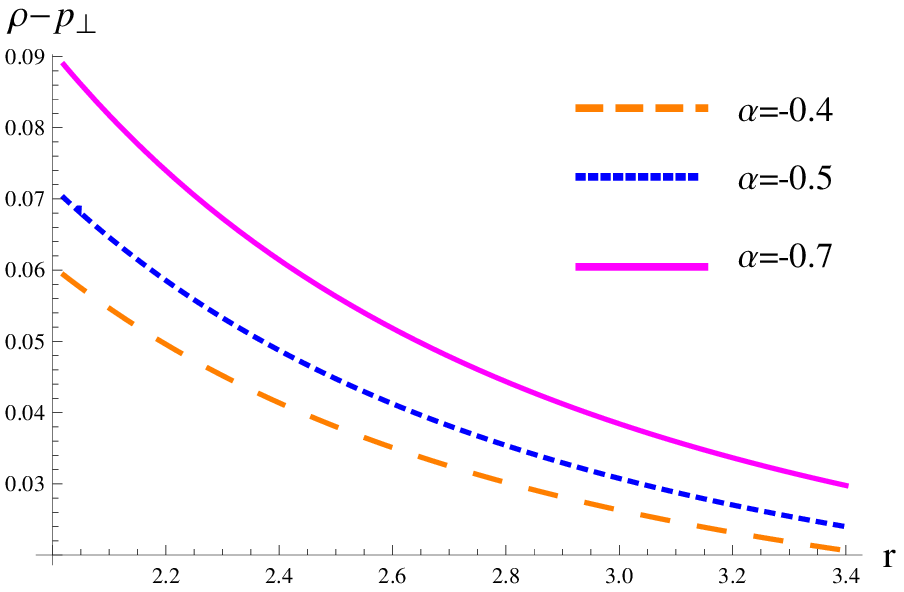,width=0.35\linewidth} \caption{Energy
conditions for case III.}\label{411}
\end{figure}

\begin{figure}\center
\epsfig{file=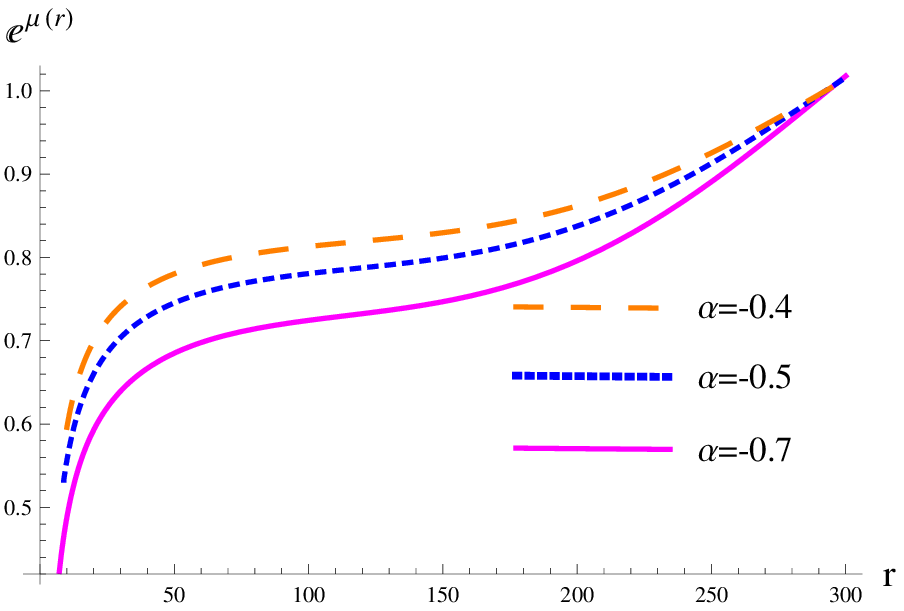,width=0.35\linewidth}\epsfig{file=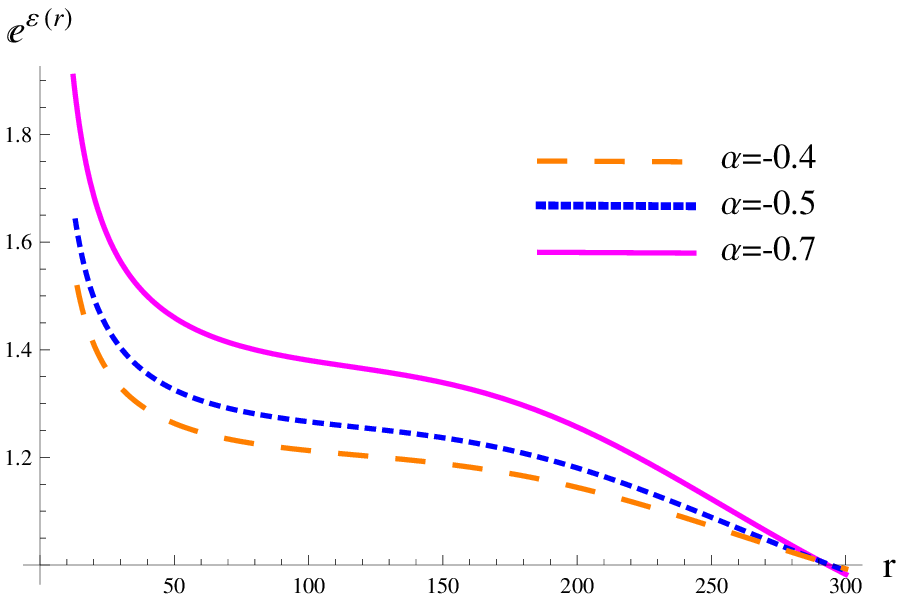,width=0.35\linewidth}
\caption{Metric potentials for case III.}\label{412}
\end{figure}

Here, we evaluate a specific solution by inserting $a_1=1.4$ and
$a_2=3$ in Eq.(\ref{42}) leading to
\begin{eqnarray}\nonumber
&&\frac{1}{r\Upsilon(r)}\left(\alpha e^{-
b(r)}\left(r\Upsilon(r)\left(r\Upsilon'(r)\left(e^{ b
(r)}h_{2}'(r)+2.2h_{1}'(r)\right)+\Upsilon(r)\left(e^{
b(r)}h_{2}'(r)\right.\right.\right.\right.\\\nonumber
&&\left.\left.\left.\left.\times \left(0.75 \alpha
rh_{1}'(r)+0.75ra'(r)+0.5\right)+1.5 rh_{1}''(r)+h_{1}'(r)\left(0.75
\alpha rh_{1}'(r)\right.\right.\right.\right.\right.\\\nonumber
&&\left.\left.\left.\left.\left.+1.5ra'(r)-0.75rb'(r)+2.9\right)\right)\right)+h_{2}(r)
e^{ b(r)}\left(r\Upsilon(r)\left(\Upsilon'(r)\left(2.2\alpha
rh_{1}'(r)\right.\right.\right.\right.\right.\\\nonumber
&&\left.\left.\left.\left.\left.+2.2r
a'(r)+3.8\right)+2r\Upsilon''(r)\right)+\Upsilon^2(r)\left(r\left(1.5
\alpha
rh_{1}''(r)+0.75\alpha^2rg'^2(r)\right.\right.\right.\right.\right.\\\nonumber
&&\left.\left.\left.\left.\left.+\alpha
h_{1}'(r)\left(1.5ra'(r)+2.9\right)+1.5ra''(r)+0.75ra'^2(r)+2.9
a'(r)\right)+0.4\right)\right.\right.\right.\\\label{45}
&&\left.\left.\left.+0.3r^2\omega
_{BD}\Upsilon'^2(r)\right)\right)\right)=0.
\end{eqnarray}
The extended solution is formulated by plugging the values of $
b(r),~a(r)$ and $h_{2}(r)$ in Eqs.(\ref{2}) and (\ref{45}) and
solving them for initial conditions of case I with $M=1$. A
singularity exists at $r=0$ since the plot of $\mathcal{K}$ in
Figure \ref{409} tends to infinity for $r\rightarrow0$. The state
variables of the solution are plotted in Figure \ref{410} for chosen
values of the decoupling parameter. The energy density and
tangential pressure are maximum at the horizon and decrease
monotonically as $r$ increases. The decrease in the values of
decoupling parameter causes an increase in density and transverse
pressure whereas the radial pressure increases with increase in
$\alpha$. The energy constraints are satisfied by
$\Theta^\gamma_\delta$ (Figure \ref{411}) ensuring the presence of
normal matter. The metric functions represent a spacetime that is
asymptotically flat as shown in Figure \ref{412}.

\section{Conclusions}

In this paper, we have focused on developing new solutions through
the EGD technique in the background of SBD theory. An additional
source is incorporated in the simple seed source such as vacuum or
isotropic fluid. The EGD approach reduces the degrees of freedom in
the system of field equations by introducing geometric deformations
in the metric functions related to the seed source. With the help of
these transformations, the original system is disintegrated into two
sets which exclusively correspond to the seed and new sources. We
have considered the seed source as a vacuum and employed the metric
potentials of Schwarzschild spacetime to specify the associated
system. The extensions of this solution in the presence of
$\Theta^\gamma_\delta$ are evaluated by imposing constraints on the
metric functions and the $\Theta$ sector. The condition
$\mu=-\varepsilon$ is implemented to ensure that Killing and causal
horizons coincide at $r=2M$. The number of unknown constants has
further been reduced by imposing the EoS
$\Theta_0^0=a_1\Theta_1^1+a_2\Theta_2^2$. Three solutions have been
generated corresponding to
\begin{itemize}
\item $a_1=-1$ and $a_2=-2$ (represents a traceless
additional source).
\item $a_1=-\frac{1}{K}$ and $a_2=0$ (provides a barotropic
fluid distribution).
\item $a_1=1.4$ and $a_2=3$ (yields a particular solution).
\end{itemize}
It is worth mentioning here that the linear EoS provides the
Schwarzschild or Kiselev BH when EGD approach is applied to
Schwarzschild metric in GR \cite{26}. The massive scalar field in
SBD gravity naturally induces anisotropy in the solution. Therefore,
metrics different than Schwarzschild are obtained in this theory.
The massive scalar field has been obtained by solving the wave
equation $V(\Upsilon))=\frac{1}{2}m_{\Upsilon}^2\Upsilon^2$ with
$m_\Upsilon=0.1$. The behavior of physical parameters of the
extended solutions has been investigated for
$\alpha=-0.4,~-0.5,~-0.7$ and $\omega_{BD}=60$.

The energy density in all cases is positive for negative values of
the decoupling parameter only which results in negative radial
pressure. This behavior of matter variables is consistent with the
work in \cite{14}. Moreover, higher energy density is observed for
lower values of $\alpha$. The nature of the extra source has also
been checked through energy conditions. The analysis of state
parameters has revealed that the extended models corresponding to
cases I and III are consistent with energy conditions. This implies
that $\Theta^\gamma_\delta$ is sourced by normal matter. However,
the extended model obtained for a barotropic EoS violates the energy
conditions indicating the presence of exotic matter. Moreover, the
asymptotic behavior of the extended solutions has been checked
through the behavior of $e^\mu$ and $e^\varepsilon$. The metrics
formulated for I and III are asymptotically flat whereas the metric
obtained for II does not approach to a flat spacetime when
$r\rightarrow\infty$. All the BH solutions mentioned above have a
singularity covered by the horizon at $r=2M$. Moreover, the extended
Schwarzschild solutions obtained through the technique of MGD in GR
violate the dominant energy condition. However, in the context of
SBD gravity two solutions have been generated that adhere to all the
energy constraints. It is interesting to mention here that all the
results of GR can be retrieved for $\Upsilon=\text{constant}$ and
$\omega_{BD}\rightarrow \infty$. 

\end{document}